\documentclass[smallextended]{svjour3}       

\usepackage{xcolor} 
\definecolor{LightGray}{gray}{0.9}
\usepackage{graphicx}
\usepackage[sort]{natbib}
\usepackage{multirow}
\usepackage{enumerate} 

\begin{document}
\title{Leveraging Structural Properties of Source Code Graphs for Just-In-Time Bug Prediction}


\author{Md Nadim         \and
        Debajyoti Mondal \and
        Chanchal K. Roy
}


\institute{University of Saskatchewan \at
              \email{mdn769@usask.ca} \\
              \email{d.mondal@usask.ca} \\
              \email{chanchal.roy@usask.ca}
}

\date{Received: date / Accepted: date}

\maketitle

\begin{abstract}
The most common use of data visualization is to minimize the complexity for proper understanding. A graph is one of the most commonly used representations for understanding relational data. It produces a simplified representation of data that is challenging to comprehend if kept in a textual format. In this study, we propose a methodology to utilize the relational properties of source code in the form of a graph to identify Just-in-Time (JIT) bug prediction in software systems during different revisions of software evolution and maintenance. We presented a method to convert the source codes of commit patches to equivalent graph representations and named it Source Code Graph (SCG). To understand and compare multiple source code graphs, we extracted several structural properties of these graphs, such as the density, number of cycles, nodes, edges, etc. We then utilized the attribute values of those SCGs to visualize and detect buggy software commits. We process more than 246K software commits from 12 subject systems in this investigation. Our investigation on these 12 open-source software projects written in C++ and Java programming languages shows that if we combine the features from SCG with conventional features used in similar studies, we will get the increased performance of Machine Learning (ML) based buggy commit detection models. We also find the increase of F1~Scores in predicting buggy and non-buggy commits statistically significant using the Wilcoxon Signed Rank Test. Since SCG-based feature values represent the style or structural properties of source code updates or changes in the software system, it suggests the importance of careful maintenance of source code style or structure for keeping a software system bug-free.


\end{abstract}

\section{Introduction}
\label{sec:introduction}
Fixing software bugs or faults is one of the most common activities in the software development and maintenance process, which is estimated to take about 50-80\% of developers' time [\cite{maintenance-time-stat}]. Fixing a detected bug may also be a reason for inducing new bugs in software systems [\cite{2019:FSE:Wen:EEC:3338906.3338962}]. Delay in catching a bug when it is first induced in a software system may also make the detection and fixing process more complicated. As a result, it can reduce the reliability and acceptance of the software system. On the other hand, if we can detect an induced bug early, it could significantly reduce the impacts because of the availability of resources such as developers (who change over time). 

Existing studies to detect and investigate the presence of a bug in software commits either use statistical metadata [\cite{CommitGuruRosen, 2019:Borg:SZZUnleashed}] of those commits extracted from its subversion (SVN) maintenance system or repositories such as GitHub. Some studies also combine those statistical data and natural language-based textual processing of source code [\cite{ReducingFeatures-BIC}] based features. However, to the best of our knowledge, the effect of coding structure or patterns on the presence of a bug in software systems is not adequately investigated in existing studies. The M.Sc. thesis [\cite{nadim-thesis-msc}] done by the first author of this manuscript reported an investigation of the patterns of source code in commit patch, which show improvement in detecting bug inducing commits from six open-source software projects. It performed an inquiry into 642 manually labelled datasets. Although the findings of that study showed that the use of code pattern-based feature values in ML-based BIC detection increases the bug detection performance measures, the datasets are from six manually labelled software systems, which are not much larger considering the count of data instances in a real word software environment. In this study, we performed an investigation on the data instances of 12 open-source C++ and Java-based software projects, and our total data instances are 246,279. We modelled patterns in the source code as graphs, referred to as the Source Code Graphs (SCG). We investigated the possibility of using the attributed values of SCG as feature values to visualize and detect buggy commits in software systems. We show the summary of our datasets in Table \ref{tab:vis-subject-systems} from the 12 subject systems. As the number of subject systems in this investigation is quite extensive, the results of this study will provide robust evidence about the generalizability of the findings.

The motivation for this investigation is to find practical evidence about different coding styles that may considerably impact the overall quality of software systems. If we can find specific styles or patterns, we can produce a programming style guidance for a developer to make the software development, evolution \& maintenance activities more effective. In this study, we want to investigate the possibility of identifying bug presence in software commits using the feature values extracted from the visual representation of source code. We collected two types of data labels and initial datasets with 15 feature values of 12 open-source software projects from the implementation of Commit Guru [\cite{CommitGuruRosen}], which can mine risky commits from any GIT SCM\footnote{https://git-scm.com/} source code manager repository. They initially clone project data from its GitHub repository and then process the commit logs for extracting 13 change-level metric values. They also describe their approach in another study [\cite{Kamei:2013:LES:2498737.2498844}]. They also utilize SCM blame/annotate\footnote{https://docs.github.com/en/repositories/working-with-files/using-files/tracking-changes-in-a-file} functionality for processing a repository and produce a dataset containing different commit categories and bug presence information. Several studies [\cite{NaturalnessBug, LargeChangeReason, WhatChangesinWhere}] also applied a similar approach for detecting a commit which is fixing a bug and then tracking the commit, which introduced the fixed bug in the software project. The bug introducing commits are generally considered as risky or buggy commits. \citet{NaturalnessBug} also reported that the technique used in these studies to identify bug-fixing commits is 96\% accurate. We are introducing two new categories of Source Code Graphs (SCG)-based feature values in this study. As we wanted to compare the performance of SCG-based features and the features commonly used in similar studies in detecting buggy and non-buggy commits, we find the initial dataset prepared by Commit Guru appropriate to take as a starter.  We can also argue that these starter datasets we use are accurate and reasonably extensive enough for our investigation, considering the popularity of the applied approach in this research area.

We added 24 new feature values we extracted from these 12 software projects to perform this investigation. Our feature values represent the coding style or structure of the software developer, and we named them Source Code Graph (SCG) based features. The 15 feature values used in the study [\cite{CommitGuruRosen}] are most common in similar machine learning-based software quality detection models. We can extract most of these feature values, such as the number of lines added/ deleted/ modified in a commit, density of change, developer age/ time from the GitHub repository of any software project. In this study, we named these GitHub statistics-based Commit Guru features and used the term `C' features to refer to them in this manuscript. CommitGuru also provided two types of data labelling, one as the presence of a bug that is either True or False. The other is the classification of commits, representing different categories (e.g., merge, perfective, preventive, etc.).

The primary motivation behind this study is to see whether the network complexity of source code graphs (SCGs) can detect the buggy commits and whether this technique outperforms the usual feature-based ML approaches. We extracted 12 Source Code Graph (SCG) based features from both the added (represented as A) and deleted (represented as D) code fragments from the commits of all the 12 software projects. Therefore, we have 12 features from added commit lines and 12 from deleted commit lines, making $12 \times 2 = 24$ features representing the Source Code Graph (SCG) of commit patches.  We listed the names of SCG based features in Table \ref{tab:graph-features}. We selected features inspired by the   `node embedding' in graph neural networks~[\cite{DBLP:journals/debu/HamiltonYL17}] and metrics for graph comparison~[\cite{gcomp}]. We wanted to detect buggy software commits based on the source code structures of added and deleted lines in software commits. So, we investigated the graph-based features as they have been successfully used to represent and compare graphs structure in graph neural networks.

We also show the mean and standard deviation of SCG-based feature values in buggy and non-buggy commits in Table \ref{tab:graph-features}. This table calculates the mean and standard deviation from the feature values of all the 12 subject systems combined. We can observe a considerable difference in values between buggy and non-buggy commits for both mean and standard deviation in the table. We also performed a t-Test [\cite{student1908tTest}] on the data samples in each feature value individually from Commit Guru and Source Code Graphs to see whether the difference in mean values of various features in buggy commits and non-buggy commits are statistically significant or not. We found the difference of mean values in most of the buggy and non-buggy commits' features in both Commit Guru (our baseline 'C') and our proposed SCG (A and D)  are statistically significant. However, the means from a few features are not statistically significant in some of the 12 subject systems; among them, the majority have come from the Commit Guru (C) features. These scenarios motivate us to investigate buggy and non-buggy commit detection performance combined with Commit Guru and SCG-based features using machine learning detection models. 

\begin{table}
\centering
\caption{\label{tab:vis-subject-systems}\textsc{The 12 subject systems Used for the Study}}
\begin{tabular}{|c|l|l|c|} 
\hline
\textbf{SL. No} & \multicolumn{1}{c|}{\textbf{Subject~Systems}} & \textbf{Prog.~Lang.} & \textbf{Commit~Instances}  \\ 
\hline
1               & Accumulo                                      & Java                 & 11,045                                      \\ 
\hline
2               & Ambari                                        & Java                 & 24,588                                      \\ 
\hline
3               & Bitcoin                                       & CPP                  & 16,759                                      \\ 
\hline
4               & Camel                                         & Java                 & 26,911                                      \\ 
\hline
5               & Jackrabbit                                    & Java                 & 8,909                                       \\ 
\hline
6               & Jenkins                                       & Java                 & 28,923                                      \\ 
\hline
7               & Litecoin                                      & CPP                  & 17,836                                      \\ 
\hline
8               & Lucene                                        & Java                 & 31,957                                      \\ 
\hline
9               & Mongo                                         & CPP                  & 50,303                                      \\ 
\hline
10              & Oozie                                         & Java                 & 2,280                                       \\ 
\hline
11              & Tomcat                                        & Java                 & 23,153                                      \\ 
\hline
12              & Zxing                                         & Java                 & 3,615                                       \\ 
\hline
\multicolumn{3}{c}{\textbf{Total Number of Commit Instances:}}                         & \multicolumn{1}{c}{\textbf{246,279}}   \\
\hline
\end{tabular}
\end{table}

\begin{table}
\caption{\label{tab:graph-features}\textsc{Source Code Graph (SCG) based Feature Names \& their Comparison of Mean and Standard Deviation in Buggy and Non-buggy Commits. We combine Add (A) and Delete (D) types of SCG based features while calculating the Mean and STDEV.}}
\centering
\begin{tabular}{|c|l|c|c|c|c|} 
\hline
\textbf{SL. No.~} & \multicolumn{1}{c|}{\begin{tabular}[c]{@{}c@{}}\textbf{SCG-based}\\\textbf{~Feature Names}\end{tabular}} & \begin{tabular}[c]{@{}c@{}}\textbf{Mean}\\\textbf{~(Buggy)}\end{tabular} & \begin{tabular}[c]{@{}c@{}}\textbf{Mean}\\\textbf{~(Non-Buggy)}\end{tabular} & \begin{tabular}[c]{@{}c@{}}\textbf{STDEV}\\\textbf{~(Buggy)}\end{tabular} & \begin{tabular}[c]{@{}c@{}}\textbf{STDEV}\\\textbf{~(Non-Buggy)}\end{tabular}  \\ 
\hline
1                 & Number
  of Cycles~                                                                                                                   & 56.05                                                                    & 13.81                                                                        & 33.9                                                                      & 9.76                                                                           \\ 
\hline
2                 & Graph
  Density~                                                                                                                      & 0.04                                                                     & 0.05                                                                         & 0.04                                                                      & 0.05                                                                           \\ 
\hline
3                 & Number
  of Nodes~                                                                                                                    & 23.05                                                                    & 14.14                                                                        & 5.51                                                                      & 5.97                                                                           \\ 
\hline
4                 & Number
  of Edges~                                                                                                                    & 43.92                                                                    & 23.76                                                                        & 13.08                                                                     & 11.23                                                                          \\ 
\hline
5                 & Average
  In Degree~                                                                                                                  & 1.62                                                                     & 1.24                                                                         & 0.33                                                                      & 0.48                                                                           \\ 
\hline
6                 & Average
  Out Degree~                                                                                                                 & 1.62                                                                     & 1.24                                                                         & 0.33                                                                      & 0.48                                                                           \\ 
\hline
7                 & Maximum
  Degree~                                                                                                                     & 12.3                                                                     & 8.22                                                                         & 3.59                                                                      & 3.9                                                                            \\ 
\hline
8                 & Minimum
  Degree~                                                                                                                     & 0.95                                                                     & 0.82                                                                         & 0.17                                                                      & 0.28                                                                           \\ 
\hline
9                 & Sum
  of Degree~                                                                                                                      & 87.84                                                                    & 47.52                                                                        & 26.17                                                                     & 22.47                                                                          \\ 
\hline
10                & Average
  Degree~                                                                                                                     & 3.15                                                                     & 2.46                                                                         & 0.85                                                                      & 1.16                                                                           \\ 
\hline
11                & Median
  Degree~                                                                                                                      & 2.47                                                                     & 1.96                                                                         & 0.53                                                                      & 0.81                                                                           \\ 
\hline
12                & Number
  of Self Loops~                                                                                                               & 0.99                                                                     & 0.77                                                                         & 0.19                                                                      & 0.31                                                                           \\
\hline
\end{tabular}
\end{table}
\begin{table}
\caption{\label{tab:vis-commit-classification}\textsc{Number of Data Instances and Its Assigned Class Labels (0--6) in different commit categories}}
\addtolength{\tabcolsep}{-1pt}
\centering
\begin{tabular}{|l|c|c|c|c|c|c|c|} 
\hline
\multicolumn{1}{|c|}{\begin{tabular}[c]{@{}c@{}}\textbf{Subject}\\\textbf{System}\end{tabular}} & \begin{tabular}[c]{@{}c@{}}\textbf{None}\\\textbf{(0)}\end{tabular} & \begin{tabular}[c]{@{}c@{}}\textbf{Merge}\\\textbf{~(1)}\end{tabular} & \begin{tabular}[c]{@{}c@{}}\textbf{Corrective}\\\textbf{(2)}\end{tabular} & \begin{tabular}[c]{@{}c@{}}\textbf{Preventive}\\\textbf{~(3)}\end{tabular} & \begin{tabular}[c]{@{}c@{}}\textbf{Feature}\\\textbf{Addition (4)}\end{tabular} & \begin{tabular}[c]{@{}c@{}}\textbf{Non}\\\textbf{Functional (5)}\end{tabular} & \begin{tabular}[c]{@{}c@{}}\textbf{Perfective}\\\textbf{~(6)}\end{tabular}  \\ 
\hline
Accumulo                                                                                        & 2,840                                                               & 3,409                                                                 & 2,192                                                                 & 354                                                                    & 1,534                                                                      & 450                                                                       & 266                                                                    \\ 
\hline
Ambari                                                                                          & 13,602                                                              & 495                                                                   & 5,184                                                                 & 483                                                                    & 4,015                                                                      & 209                                                                       & 600                                                                    \\ 
\hline
Bitcoin                                                                                         & 3,371                                                               & 4,598                                                                 & 3,127                                                                 & 1,798                                                                  & 3,197                                                                      & 348                                                                       & 320                                                                    \\ 
\hline
Camel                                                                                           & 10,315                                                              & 196                                                                   & 8,492                                                                 & 1,394                                                                  & 4,645                                                                      & 964                                                                       & 905                                                                    \\ 
\hline
Jackrabbit                                                                                      & 4,269                                                               & 3                                                                     & 1,611                                                                 & 571                                                                    & 1,797                                                                      & 446                                                                       & 212                                                                    \\ 
\hline
Jenkins                                                                                         & 11,704                                                              & 4,372                                                                 & 6,874                                                                 & 859                                                                    & 3,756                                                                      & 873                                                                       & 485                                                                    \\ 
\hline
Litecoin                                                                                        & 4,307                                                               & 6,346                                                                 & 2,843                                                                 & 658                                                                    & 3,083                                                                      & 330                                                                       & 269                                                                    \\ 
\hline
Lucene                                                                                          & 11,328                                                              & 614                                                                   & 8,474                                                                 & 2,329                                                                  & 6,292                                                                      & 2,024                                                                     & 896                                                                    \\ 
\hline
Mongo                                                                                           & 23,577                                                              & 2,279                                                                 & 9,034                                                                 & 4,285                                                                  & 8,238                                                                      & 842                                                                       & 2,048                                                                  \\ 
\hline
Oozie                                                                                           & 1,192                                                               & 1                                                                     & 482                                                                   & 124                                                                    & 350                                                                        & 95                                                                        & 36                                                                     \\ 
\hline
Tomcat                                                                                          & 8,302                                                               & 40                                                                    & 8,897                                                                 & 660                                                                    & 3,537                                                                      & 574                                                                       & 1,143                                                                  \\ 
\hline
Zxing                                                                                           & 1,785                                                               & 75                                                                    & 813                                                                   & 125                                                                    & 653                                                                        & 37                                                                        & 127                                                                    \\
\hline
\end{tabular}
\end{table}

\begin{table}
\caption{\label{tab:vis-bug-presence}\textsc{Number of data instances and its assigned class labels (0--1) based on bug presence}}
\centering
\begin{tabular}{|l|c|c|} 
\hline
\textbf{Subject Systems~}            & \textbf{Buggy Commits (L = 1)}          & \textbf{Non Buggy Commits (L = 0)~}       \\ 
\hline
\textbf{Accumulo}                    & 1,763                               & 9,282                                 \\ 
\hline
\textbf{Ambari}                      & 5,502                               & 19,086                                \\ 
\hline
\textbf{Bitcoin}                     & 3,318                               & 13,441                                \\ 
\hline
\textbf{Camel}                       & 9,834                               & 17,077                                \\ 
\hline
\textbf{Jackrabbit}                  & 1,771                               & 7,138                                 \\ 
\hline
\textbf{Jenkins}                     & 4,640                               & 24,283                                \\ 
\hline
\textbf{Litecoin}                    & 2,867                               & 14,969                                \\ 
\hline
\textbf{Lucene}                      & 8,011                               & 23,946                                \\ 
\hline
\textbf{Mongo}                       & 12,588                              & 37,715                                \\ 
\hline
\textbf{Oozie}                       & 453                                 & 1,827                                 \\ 
\hline
\textbf{Tomcat}                      & 5,677                               & 17,476                                \\ 
\hline
\textbf{Zxing}                       & 847                                 & 2,768                                 \\ 
\hline
\multicolumn{1}{r}{\textbf{Total:~}} & \multicolumn{1}{c}{\textbf{57,271}} & \multicolumn{1}{c}{\textbf{189,008}} \\
\hline
\end{tabular}
\end{table}

\begin{figure}
\includegraphics[width=\textwidth]{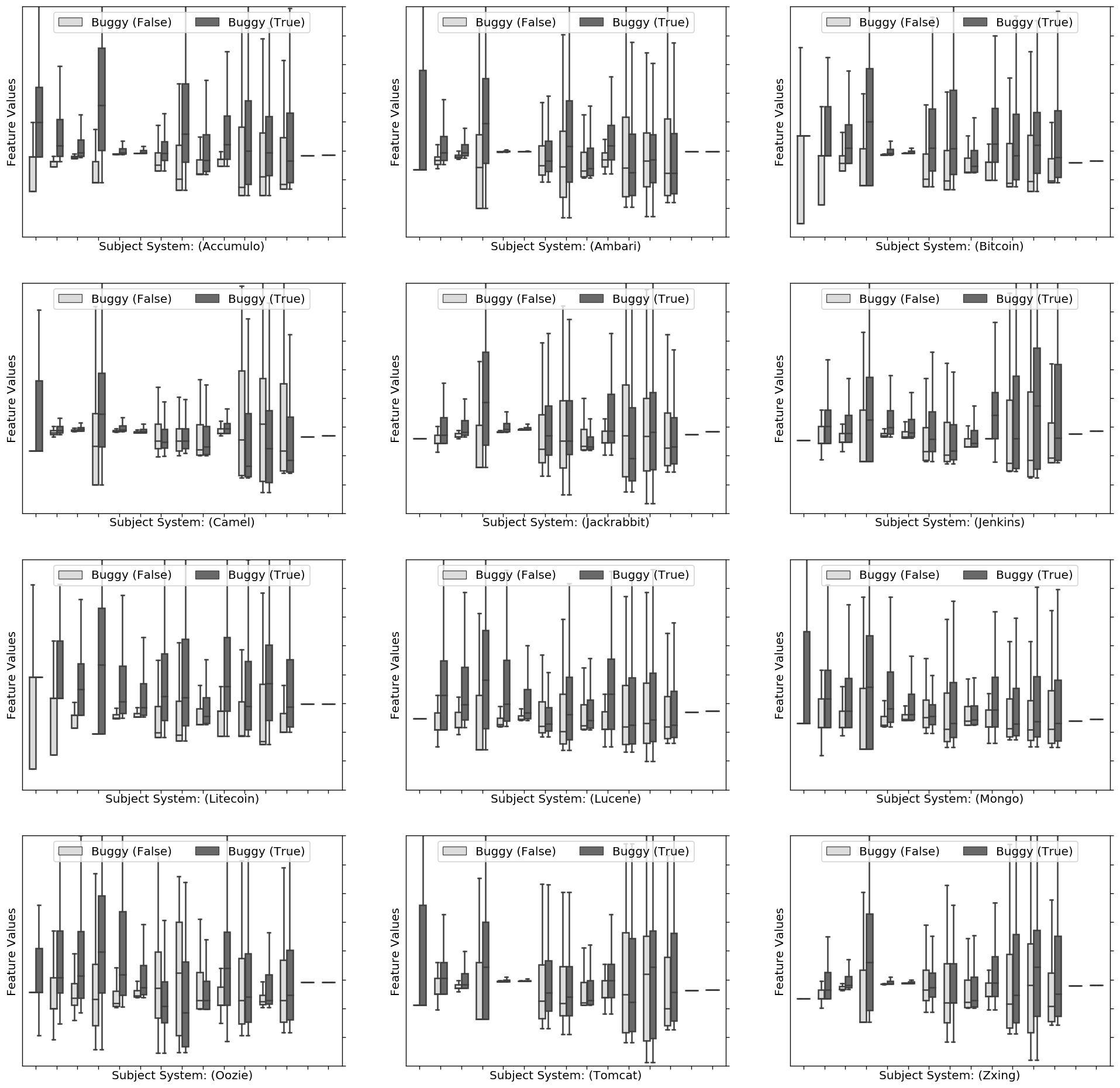}
\caption{Commit Guru (C) feature distribution in all the Subject Systems. We can see an almost overlapping distribution of feature values from the buggy and non-buggy commit labels in all the 12 subject systems.}
\label{fig:CFeaturesBox}       
\end{figure}

\begin{figure}
\includegraphics[width=\textwidth]{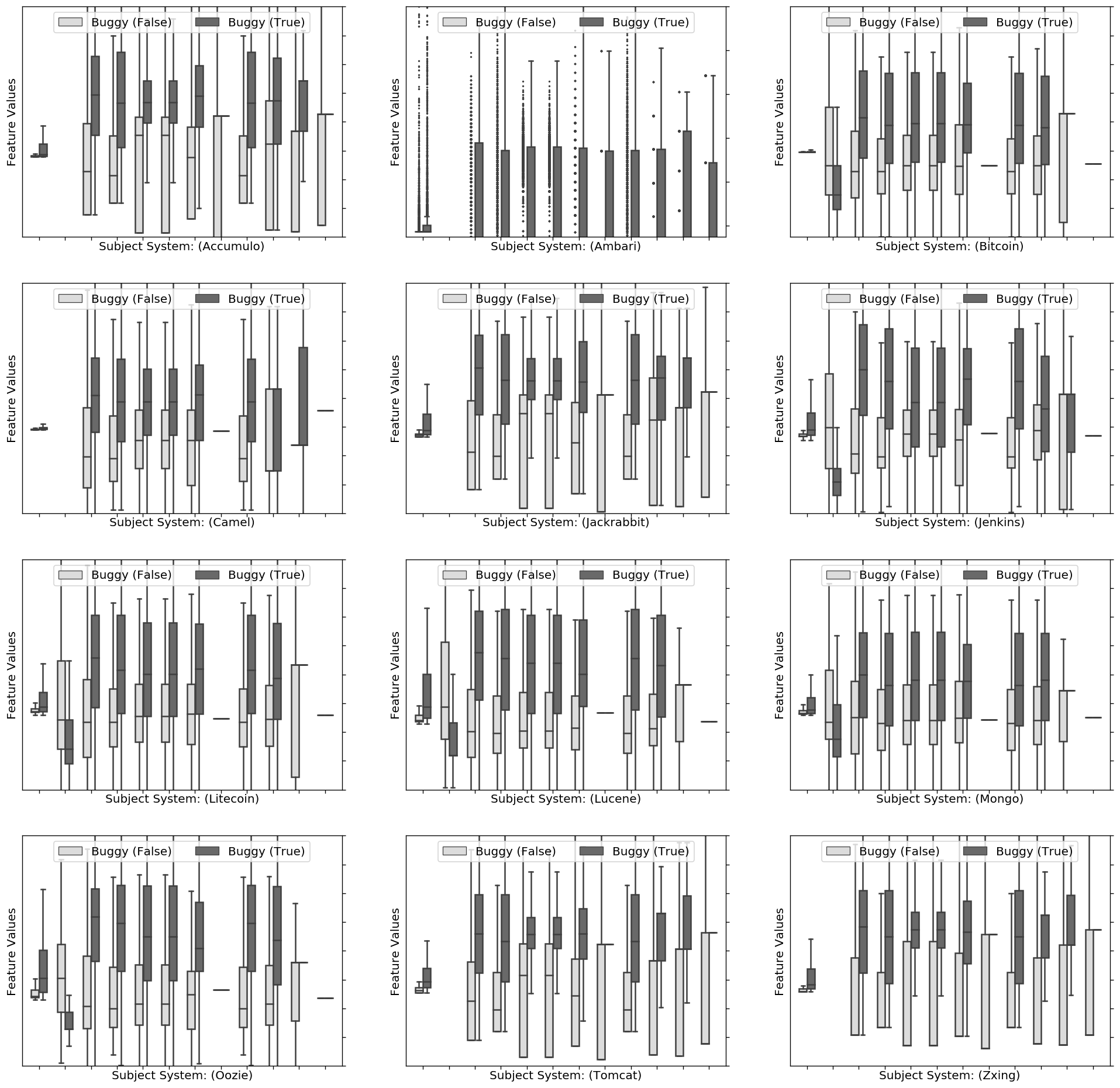}
\caption{SCG Add (A) feature distribution in all the Subject Systems. We can see a large portion of the distribution of feature values from the buggy and non-buggy commit labels are not overlapping in all the 12 subject systems.}
\label{fig:SCGFeaturesADD}       
\end{figure}

\begin{figure}
\includegraphics[width=\textwidth]{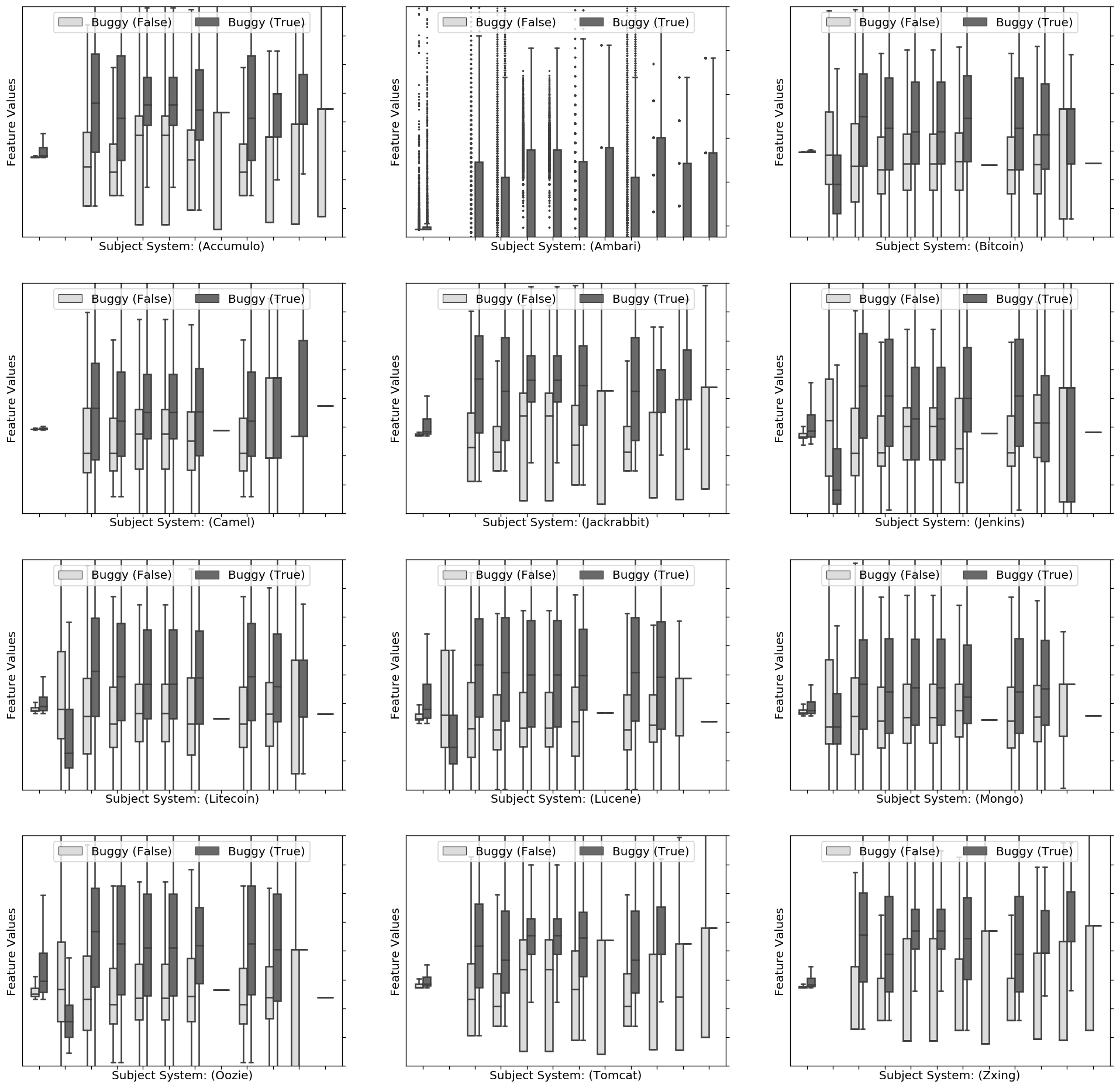}
\caption{SCG Delete (D) feature distribution in all the Subject Systems. Similar to the distribution of SCG Add (A), we can see that a large portion of the distribution of feature values from the buggy and non-buggy commit labels does not overlap in all the 12 subject systems.}
\label{fig:SCGFeaturesDEL}       
\end{figure}

We have computed some standardized [\cite{standardization}] box plots of all Commmit Guru and SCG features in Figures \ref{fig:CFeaturesBox}, \ref{fig:SCGFeaturesADD}, and \ref{fig:SCGFeaturesDEL} to see the distribution of the feature values in buggy and non-buggy commits and compare them with each other. The standardization process made all the feature values easier to plot on the same scale. To standardize, we first subtract the mean value from each feature value and then divide it by standard deviation. After the standardization, we found all the SCG feature values on the scale of $-2.50$ to $+2.50$. After that, we draw the box plots for all the subject systems. In the distribution of Commit Guru (C) in Figure \ref{fig:CFeaturesBox}, we can see that a higher portion of the feature values from buggy and non-buggy commits overlap each other. On the other hand, the distribution of SCG based Add (A) and Delete (D) feature values in Figures \ref{fig:SCGFeaturesADD} \& \ref{fig:SCGFeaturesDEL}, we see most of the non-overlapping distributions of feature values from the commits when partitioned into the buggy and the non-buggy categories. For example, in all of these distributions of SCG feature values,  we found that graph $density$ in Buggy (True) is lower than the Buggy (False) commits. We also observed, most of the SCG feature values, such as number of nodes, number of edges, incoming degree, etc., are higher in buggy commits than the non-buggy commits. We can see an exception in the distribution of SCG feature values in the subject system `Ambari'. In this subject system, feature values in both the SCG A and D commit categories are distributed too far from each other in non-buggy commits than buggy commits. Although this is an exception in one subject system out of 12 in this study, it also provides evidence that SCG feature values' distribution differs in buggy and non-buggy commits. These scenarios also motivate us to investigate the capability of detecting bugs by utilizing the visual representation of source code. Findings from this study could lead us to develop autonomous bug detection systems based on the structural attributes of source code. 

We organized this study in the following sections. Section \ref{sec:back:terminology} discuss some background terminology used in this study, research hypothesis and research questions are discussed in section \ref{sec:hypo-rqs}, overall research methodology  is in section \ref{sec:methodology}, we discuss experimental results in section \ref{sec:results}, some threats to validity is discussed in section \ref{threats-validity}, related works are in section \ref{related-work}, and finally we concluded this paper in section \ref{conclusion-future-work}. 

\vspace{0.40cm}
\section{Background Terminologies}
\label{sec:back:terminology}
This section will discuss a few relevant terminologies that we have used in this paper.

\textbf{Software Evolution:} Software evolution is a continuous process to keep a software system consistent within its life cycle. The changing requirements, technologies, and stakeholder knowledge [\cite{SoftwareEvolution}] are the main factors for software evolution to keep a software system up to date. 

\textbf{Commit Operation:} The contribution of different software developers in software repositories is known as commit. Any of these software commit operations may happen for different reasons, such as fixing a bug, adding new software features, or cleaning the code [\cite{RhetoricSmallChange, LargeCommitTellUs, LargeChangeReason}]. Table \ref{tab:vis-commit-classification} identifies seven different categories of commits in our subject systems. A commit operation might be a reason for introducing a new bug in the software systems [\cite{Kamei:2013:LES:2498737.2498844, Kim:Classify:Clean:Buggy, Sliwerski-2005-induce-fixes-journal, Yin:2011:FBB:2025113.2025121, Has-bug-really-fixed}], these commits are known as Bug Inducing Commit (BIC) [\cite{nadim-thesis-msc}].  

\textbf{Commit Attributes:} These are the attributes, which could be extracted by analyzing the commits from a software repository. These attributes explain the nature of a commit operation, such as the number of changed lines (Added or Deleted), the number of changed files, changed line density (ratio of changed lines vs. the total number of lines), etc.


\textbf{t-SNE:} t-Distributed Stochastic Neighbor Embedding (t-SNE) is a technique that visualizes high dimensional datasets by providing each data point location in lower dimension (i.e. two or three) [\cite{JMLR:tSNE, tSNE}]. This technique is a state-of-the-art technique for dimensionality reduction and is well suited for data visualization. It is a variation of Stochastic Neighbor Embedding [\cite{Hinton:Roweis:2003}] that provides a much easier optimization during the reduction of dimensionality in datasets. 

The t-SNE is obtained by defining a probability distribution over pairs of high-dimensional data points such that similar data points obtain higher probabilities and dissimilar data points obtain lower probabilities. A similar probability distribution is also defined in the low-dimensional space. Then by minimizing the  Kullback-Leibler divergence between these two distributions, one can obtain a low dimensional plot (t-SNE plot) where the distances among data points in the plot attain the relative distances in the high-dimensions.

Formally,  t-SNE first calculates the probabilities $p_{i|j}$ that is proportional to the similarity of objects $x_{i}$ and $x_{j}$, as follows.
\[ p_{i|j} = \frac{exp(-||x_{i}-x_{j}||^{2}/2\sigma_i^2)}{\sum_{k\neq1}exp(-||x_{i}-x_{k}||^{2}/2\sigma_i^2)}\]

Where $\sigma_{i}$ denotes the bandwidth of the Gaussian kernels, which is set based on data density. Let $y_i$ and $y_j$ the low dimensional representations to be determined for $x_i$ and $x_j$. Then the similarities $q_{i|j}$ between $y_i$ and $y_j$ is defined as follows.  
\[ q_{i|j} = \frac{(1+||y_{i}-y_{j}||^{2})^{-1}}{\sum_{k}\sum_{l\neq k}(1+||y_{k}-y_{l}||^{2})^{-1}}\]. 

Finally, the low dimensional representations  $y_{i}$ are set by minimizing the Kullback-Leibler divergence between these two probability distributions. 

The plots generated by t-SNE often reveal clusters. We have used t-SNE to visualize the different categories of commits in Figure \ref{fig:VisClassification} and the bug presence (buggy and non-buggy labels) in commit in Figure \ref{fig:VisContainBug}.  

\section{Research Hypothesis/ Questions}
\label{sec:hypo-rqs}
In this study, our investigation evaluates the validity of the following hypothesis ($h_1$).
\begin{itemize}
  \item $h_1$: The complexity of coding structures is positively related to the presence of bugs/defects in software systems.
\end{itemize}

The rationale behind this hypothesis is that the source code having a complex coding structure is likely to contain more bugs as it demands higher efforts from the developer during the commit operation. As we are encoding the source code of a commit-patch in a graph, its complexity is captured by different graph attributes shown in Table \ref{tab:graph-features}. The validation of the hypothesis is done by analyzing the graph properties of the buggy and non-buggy commits leveraging feature visualization and statistics. We first converted the changed source code of each of the commits from all the subject systems in a graph representation, where the code statements in the modified code are converted to a network connecting nodes and edges. The nodes are formed using the tokens of code statements, and how the control of different statements is flowing is represented by the edges. Figure \ref{fig:demo-code-pattern} demonstrates a few such code patterns as the sequence of tokens, which have been reported in the M.Sc. thesis done by [\citet{nadim-thesis-msc}]. The Master's thesis also noted that the use of those Token Patterns (TPs) as feature values could enhance the performance of Bug Inducing Commit (BIC) detection using Machine Learning (ML) based detection models. Therefore, the validation of $h_1$ could answer the following two research questions: 

\begin{itemize}
  \item $RQ_1$: Can we detect bugs/defects in software systems by utilizing the complexity of coding structures? 
  \item $RQ_2$: Can we meaningfully visualize the presence of a bug in software commits using the visual (graph) attribute values of the coding structure?
\end{itemize}

The positive answer to both the research question will validate our research hypothesis.

\section{Methodology}
\label{sec:methodology}
The overall methodology of this study is divided into the following steps. 
\subsection{Dataset Collection} 
To verify the hypothesis ($h_1$) for evaluating whether the coding style or structure co-relates with the presence of bugs, we need some labelled datasets from software projects. The labels should indicate which commits contain bugs and which are not. Our selection of software projects for this study is based on the use of software projects in recent similar studies [\cite{2019:FSE:Wen:EEC:3338906.3338962, 2019:Borg:SZZUnleashed, 10.1145/3377811.3380403, 10.1145/3345629.3345639}], and the availability of a considerable amount of labelled data from each subject system. We selected 12 open-source software systems written in C++ and Java programming languages. We collected the data labels of buggy and non-buggy commits of these 12 software projects using the web-based interface of Commit Guru\footnote{http://commit.guru/}. Therefore, the availability of candidate software projects on the commit guru server was also an important criterion for selecting them for our investigation.  \citet{CommitGuruStudy} published a study to predict the risk of software commits based on the dataset they extracted from the GitHub repository of the software systems. Their implemented tool is publicly available to mine any GIT SCM (Source Code Manager) repository. We have utilized the Commit Guru tool to initially prepare the dataset and then added 24 other Source Code Graph (SCG) based feature values to each commit ID from our graph implementation of source codes to implement this study. The summary of prepared datasets is given in Tables \ref{tab:vis-subject-systems}, \ref{tab:vis-commit-classification}, and \ref{tab:vis-bug-presence}. Table \ref{tab:vis-subject-systems} lists 12 open-source software systems written in the C++ and Java programming languages of various sizes and application domains. In Table \ref{tab:vis-commit-classification}, the number of commits in each of the five commit classifications is given, and the number of the buggy and non-buggy commits in those software systems are given in Table \ref{tab:vis-bug-presence}. 

Commit Guru provided us with the labelling of the target classes, and it also provided 15 feature values extracted from the GitHub statistics of those software systems. Table \ref{tab:sample-dataset} shows a demonstration of the dataset we obtained from Commit Guru and how we combine our SCG-based features in 12 subject systems. Data value from each commit is identified by an ID (e.g., commit hash).  Each commit is represented using three different types of feature values and two different types of feature labelling. The first three columns (C1, C2, and C15) next to ID (e.g., commit id) show a sample of three feature values (out of 15), which we obtained from the website of Commit Guru. The other six columns show the sample of feature values, which we extracted from the graph attributes. To extract the feature values from graph attributes, we first converted the source code lines, which are added in a commit and which are deleted in a commit, into separate graphs and then extracted attribute values to be used as feature values to represent the commit. Actual names of graph-based feature values are given in Table \ref{tab:graph-features}. We have two different labels from each of the commits in our datasets, i) Bug Presence: L=0 represents commits with no bug, and ii) L=1 represents the buggy commits. We show the number of buggy and non-buggy commits from each subject system in Table \ref{tab:vis-bug-presence}. The commits from various software projects are also divided into seven different commit categories, which are given a numeric value, such as 0-None, 1-Merge, 2-Corrective, etc. We present these commit categories and their number of occurrences in Table \ref{tab:vis-commit-classification}.

\begin{table}
\caption{\label{tab:sample-dataset}\textsc{Sample Dataset}}
\centering
\begin{tabular}{|c|c|c|c|c|c|c|c|c|c|c|}
\hline
\multicolumn{1}{|l|}{\multirow{2}{*}{\textbf{ID}}} & \multicolumn{3}{c|}{\textbf{\begin{tabular}[c]{@{}c@{}}Features from \\ Commit   Guru (C)\end{tabular}}} & \multicolumn{3}{c|}{\textbf{\begin{tabular}[c]{@{}c@{}}Source Code \\ Added (A)\end{tabular}}} & \multicolumn{3}{c|}{\textbf{\begin{tabular}[c]{@{}c@{}}Source Code \\ Deleted   (D)\end{tabular}}} & \multirow{2}{*}{\textbf{\begin{tabular}[c]{@{}c@{}}$Buggy=1$ \\ $Others=0$\end{tabular}}} \\ \cline{2-10}
\multicolumn{1}{|l|}{}                             & \textbf{C1}                       & \textbf{C2}                      & \textbf{C15}                      & \textbf{A1}                   & \textbf{A2}                   & \textbf{A12}                   & \textbf{D1}                     & \textbf{D2}                    & \textbf{D12}                    &                                                                                \\ \hline
1                                                  & 3                                 & 4                                & 1.68                              & 4                             & 0.12                          & 13                             & 4                               & 0.13                           & 13                              & 1                                                                              \\ \hline
2                                                  & 1                                 & 4                                & 1.95                              & 4                             & 0.17                          & 10                             & 4                               & 0.17                           & 10                              & 0                                                                              \\ \hline
3                                                  & 2                                 & 2                                & 1                                 & 7                             & 0.1                           & 18                             & 7                               & 0.1                            & 18                              & 0                                                                              \\ \hline
4                                                  & 1                                 & 1                                & 0                                 & 2                             & 0.08                          & 18                             & 2                               & 0.08                           & 18                              & 1                                                                              \\ \hline
5                                                  & 1                                 & 1                                & 0                                 & 4                             & 0.13                          & 13                             & 4                               & 0.13                           & 13                              & 0                                                                              \\ \hline
\end{tabular}
\end{table}

\subsection{Converting Source Code Patterns to Graph Representation} 
Different software developers perform their programming activities uniquely when using different functionalities of a particular programming language. For example, let us assume a Java programmer who wants to test a variable $v1$ is less than or equal to another variable $v2$ or not? To do so, s/he will write a code such as $if(v1<=v2)$ or $if(v1<(v2+1))$. The code is doing the same computation in both cases, but they are written in two different patterns or structures. The difference in coding patterns done by different software developers might make it different for the software evolution \& maintenance activities. We investigate these differences in source code patterns in the form of Source Code Graph (SCG) to identify buggy commits from software revisions. 

\begin{figure}[ht]
\centering
\includegraphics[width=\textwidth]{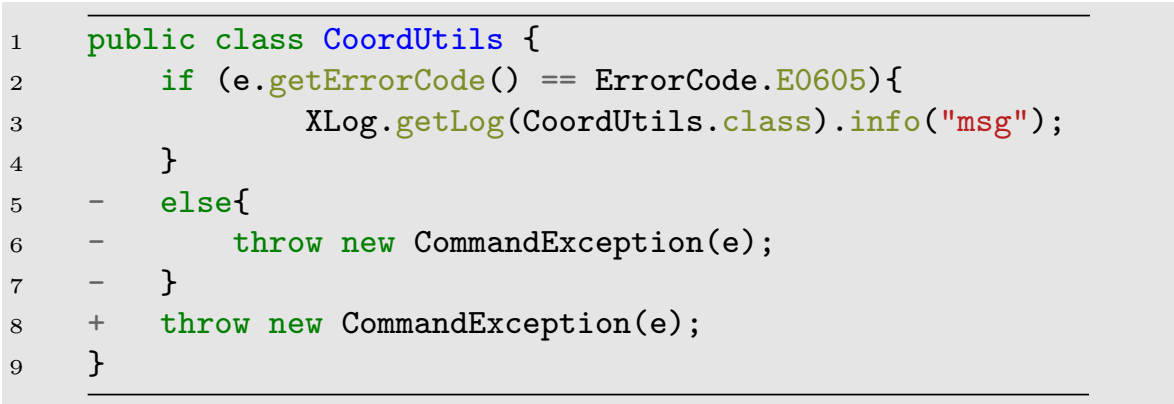}
\caption{Sample Added (+) and Deleted (-) Source Code Lines in a Commit Patch.}
\label{fig:sample-source-commit}
\end{figure}

We first extracted all the source code lines (added and deleted lines) from the commit patches to identify those patterns from the source code and saved them in separate files. We then converted those added and deleted lines into two separate graph representations. A programmer performs addition or deletion in source code lines during the commit operation of software evolution. Newly added lines are indicated by a plus sign (+), and the deleted lines are indicated by a minus sign (-) in the commit patch. We utilized these signs to identify which lines are added and which lines are deleted while processing the commit patches of our subject systems. Figure \ref{fig:sample-source-commit} shows a sample of a typical commit patch. Besides +/- signs, there are some lines without any sign. Those lines are kept to maintain the context of the commit patch. We have converted such commit patches in two different graph representations, i) taking all the added and context lines, ii) taking all the deleted and context lines. In this example, lines 5-7 are replaced by line 8, which changes the source code structure. Previously, the code implemented the \textbf{throw command exception}
under the \textbf{else-statement} of the \textbf{if-statement}, but after applying the commit patch, the programmer removed the \textbf{else-statement} and implemented the \textbf{throw command exception} independent of the \textbf{if-statement}. 

\begin{figure}
\centering
\includegraphics[width=\textwidth]{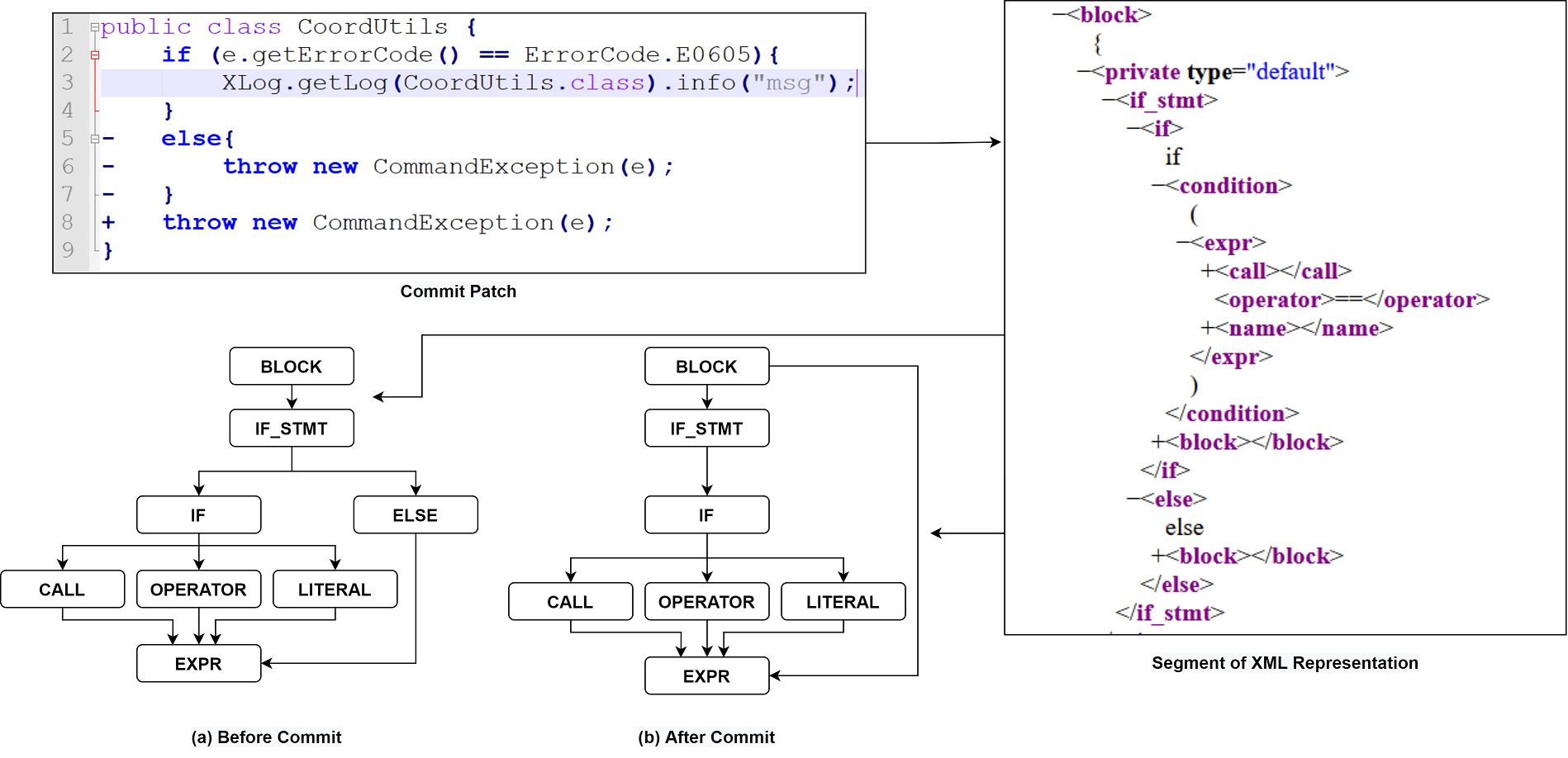}
\caption{Demonstration of source code graph (SCG) Generation}
\label{fig:genSCG}
\end{figure}

Figure \ref{fig:genSCG} shows a few more steps in generating the Source Code Graph from the commit patch shown in Figure \ref{fig:sample-source-commit}. A segment from the XML representation of source code lines shows the hierarchy of tokens converted into a graph. Two graphs in this figure show the change in code structure in that commit operation. As a developer may perform multiple changes (addition/ deletion) in a single commit, it could generate multiple graph representations in such commits. We took the union of graphs to form a single graph of a single commit for extracting graph attribute-based feature values. We use those feature values as additional features to represent commits in bug visualization and its ML-based classification to verify our research questions and hypothesis. 

\begin{figure}
\centering
\includegraphics[width=0.95\textwidth]{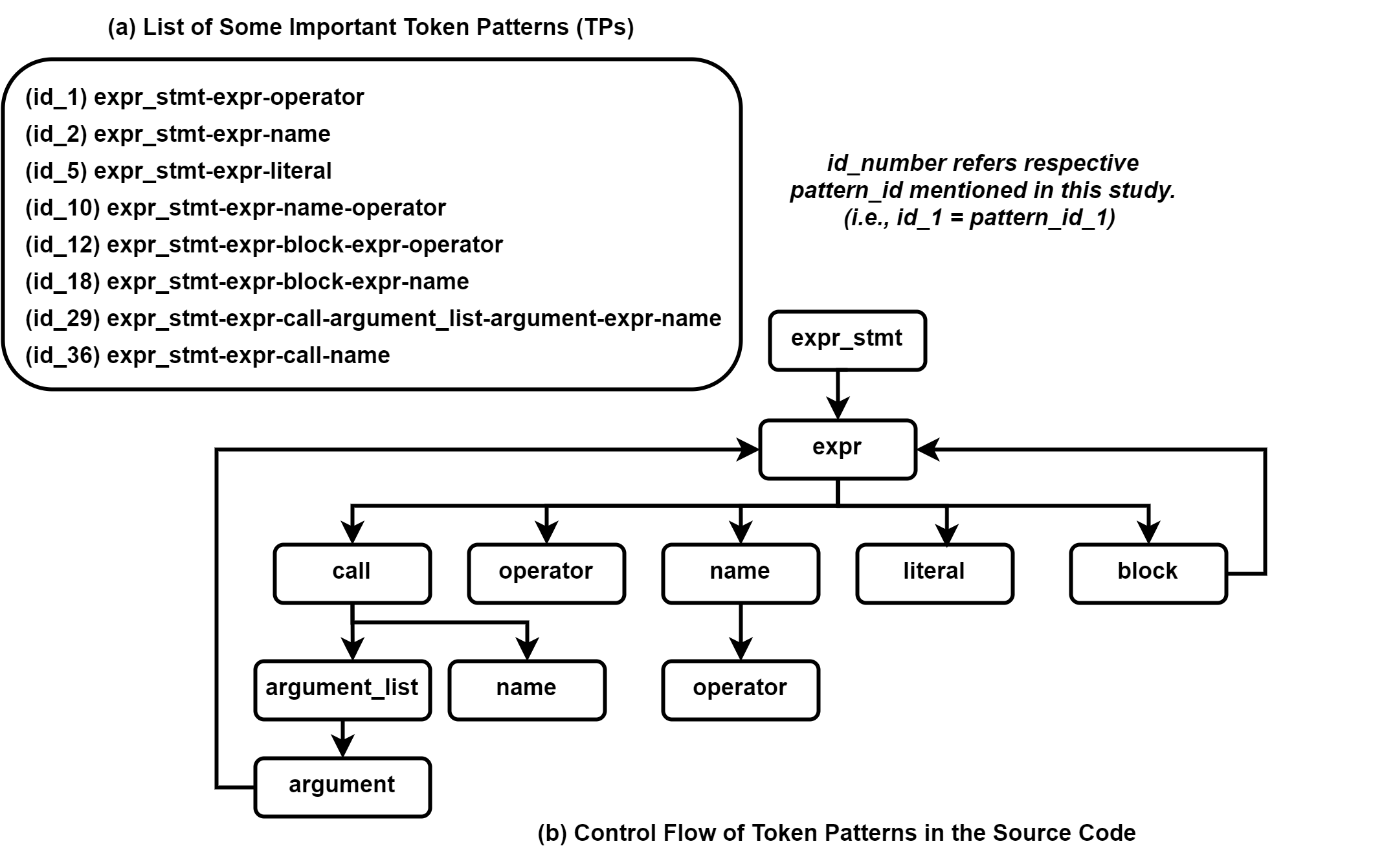}
\caption{This diagram demonstrates a few examples of extracted coding patterns, where some patterns are simple ($id\_1, id\_2$, etc.) and some have nested structures (i.e. $id\_12$ and $id\_29$, etc.). The pattern denoted by $id\_1$ represents an expression statement in the source code that ends with a single operation with an operator such as +/-. Similarly, $id\_29$ represents a nested structure that begins with an expression statement, then a function call, which has some arguments, and any of those arguments contains a nested expression. All these coding styles/patterns are extracted from the XML representation of the source files.}
\label{fig:demo-code-pattern}
\end{figure}

We then calculated its attribute values to be used as additional feature values to represent those commits. The key steps in this section are as follows. 
\begin{enumerate}[i.]
  \item Extract the source code of commit patches for each of the commits. The added and deleted lines are kept separate and processed separately. 
  \item Apply the SrcML\footnote{https://www.srcml.org/} tool on the source code to generate an XML representation of the coding pattern.
  \item Parse the XML files to extract a token structure for visualization. The hierarchy of these tokens represents the hierarchy of their occurrences in the source code (demonstrated in Figure \ref{fig:demo-code-pattern}).
  \item Generate the graphs considering the tokens as nodes, and the sequence of tokens are connected by edges. 
  \item Calculate different attribute values from the generated graphs. We extracted 12 different attribute values (Table \ref{tab:graph-features}) calculated from both of the graphs representing source code added (A) and source code deleted (D). 
  \item The attribute values are added to the dataset to be used to represent the commits.
\end{enumerate}

\subsection{Different Combinations of the Feature Values (C, A, D)} 
Extracting graph representations from the source code of commit patches provided us with two different datasets for each subject system. The feature values we used to represent commits in those datasets are listed in Table \ref{tab:graph-features}. We represented commits with those 12 graph attribute values from both the commit lines, which are added and deleted in the commit patch. Therefore, we defined each of the commits using 12 graph attribute values by added lines and another similar representation with the deleted lines. For simplicity in presentation, we will use the letter \textbf{A} and  \textbf{D} to represent results using graph-based features of added lines and deleted lines in the source code, respectively. The results using the features encoded by Commit Guru are presented by the letter \textbf{C}. We present our results using these representations of features and different combinations of them. Our results in Figures \ref{fig:ImprovementsLR}, \ref{fig:ImprovementsRF}, and \ref{fig:ImprovementsKNN} show the improvements of using SCG-based features (A, D) in the combination of GithHub Statistics based features (C) from Commit Guru implementation. For example, feature \textbf{CA} represents the result when we combine the Commit Guru and Graph (Added) datasets, and \textbf{CAD} corresponds to the combination of all three types of datasets. We have three (\textbf{C, A, D}) different kinds of features. Therefore, we can obtain seven different combinations of features (\textbf{C, A, D, CA, CD, AD, CAD}). We apply each of these seven feature combinations to detect buggy commits for all the 12 software projects. Then we find which feature combination is performing well in detecting buggy commits. We evaluated and compared those results of different feature combinations to find those which improve the buggy commit detection F1~scores. We reported and presented these results to verify our hypothesis of this study. 

\subsection{Visualizing the Datasets Utilizing t-SNE} 
\begin{figure}
\centering
\includegraphics[width=\textwidth]{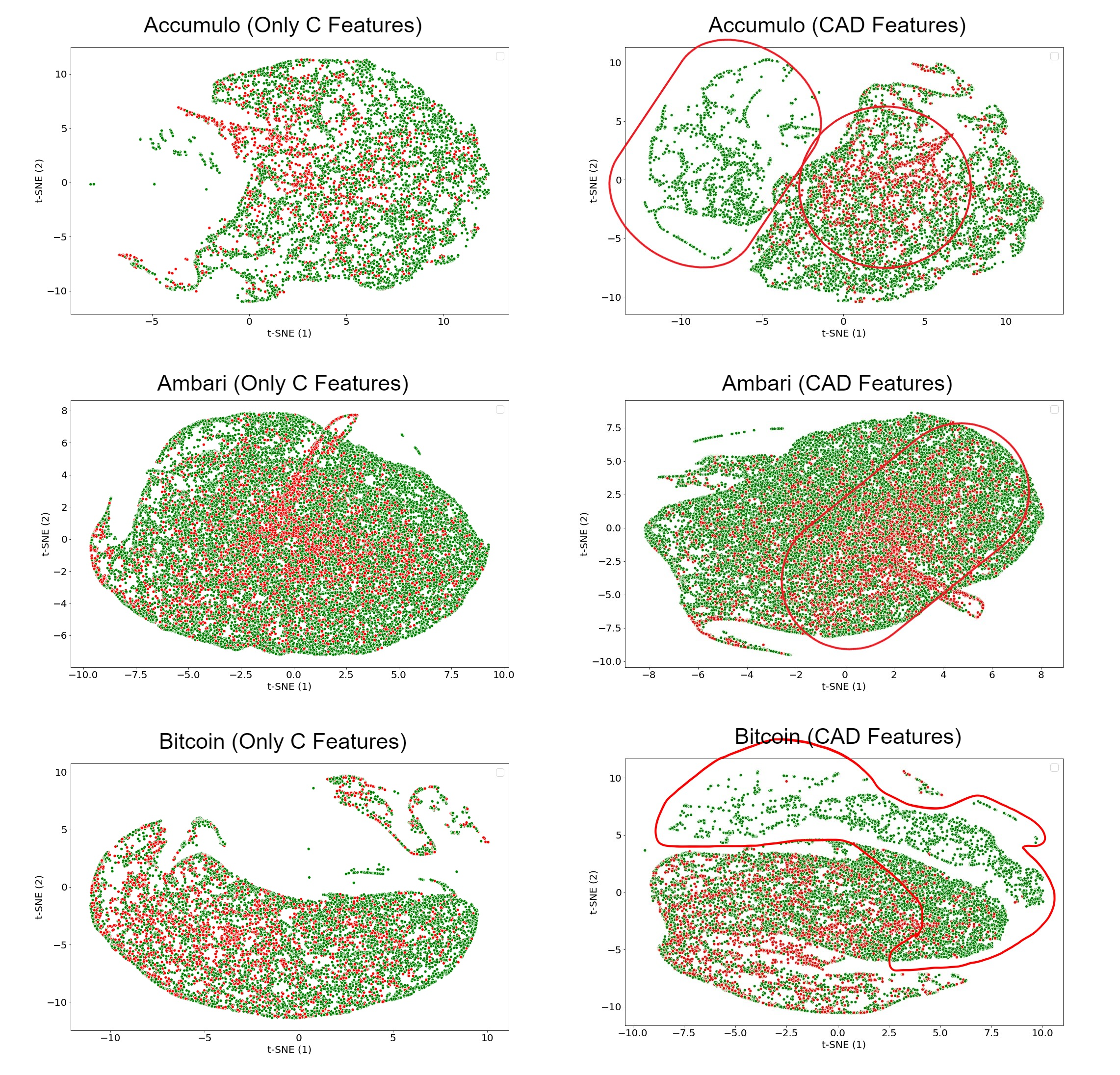}
\caption{Visualizing Commits Based on the Bug Presence using the t-SNE method. Here, red points represent the buggy commits, and green points represent non-buggy commits. We can visualize improved clusters (enclosed by red lines) of red or green points in the right where we use a combination of C, A, and D features.}
\label{fig:VisContainBug}
\end{figure}

We first evaluated the hypothesis $h_0$: Different commit labels (e.g., buggy and non-buggy) have different coding structures; by producing a visualization in Figure \ref{fig:VisContainBug}. We utilized the labelling of the commits done by Commit Guru. Six visualizations from three subject systems (selected alphabetically) are shown in this figure. It demonstrates the effectiveness of adding source code graph (SCG) based features in visualizing buggy and non-buggy commits in the source code of software projects. Three of these visualizations (in the left) are computed using only the Commit Guru or C based feature values, and the other three (in the right) have utilized the combination of C and SCG-based feature values (CAD). These visualizations are done using the t-SNE [\cite{tSNE, JMLR:tSNE}] graph visualization technique, which implements dimensionality reduction technique to plot higher dimensional data into lower dimension (2D). The red and green dots in the figure represent buggy and non-buggy commits, respectively. As we have three separate feature sets (C, A, D), we can obtain seven different feature combinations (C, A, D, CA, CD, AD, CAD) of them. We visualized all the datasets from our 12 subject systems using those seven feature combinations. Therefore, to visualize the buggy and non-buggy commits, we implemented a total of 12$\times$7=84 visualizations. Among them, six are shown in Figure \ref{fig:VisContainBug}. In this figure, \textbf{C} represents the use of feature values obtained from the Commit Guru web page, and \textbf{CAD} represents we combined Graph (Added and Deleted) features values with the Commit Guru feature values. We can see some differences in the regions enclosed by red lines in these visualizations. For example, the visualizations on the right-hand side of the figure can separately identify some regions where red or green pixels are denser/ clustered than the other regions. We also observed a similar visualization trend in our other visualizations. All of them are publicly available on our GitHub repository\footnote{https://github.com/mnadims/pubDemo/tree/main/BugInCommit}. The inclusion of SCG-based feature values also improves the clustering of the different commits categories shown in Table \ref{tab:vis-commit-classification}; however, more study is required to enhance these clusterings.

\subsection{Applying the ML-based Classification Models} 
We applied three ML-based classification models to see whether the combinations of Graph-based features and the default Commit Guru based feature can improve the detection performance of buggy and non-buggy commits. We apply machine learning models using time-sensitive detection techniques, where training data must be from past timestamps than the testing data. \citet{Tan:2015:ODP:ICSE} show that cross-validation can provide false higher precision in predicting a data instance sensitive to timestamp. A situation may happen that train the ML model with the data from the future and test it using the data from the past. We took 70\%  of the data for training the models and  30\% for testing. 

We use the configurations of Machine Learning (ML) classifiers listed in Table \ref{tab:ml-configurations} to apply the buggy, and non-buggy commit detection approaches. All the other parameters which are not mentioned in the table are set to their default values. We use the same parameter values of each ML classifier to detect buggy and non-buggy commits from all the 12 subject systems. As our goal is to compare the performance of different feature combinations (C, A, and D), any other parameter values of ML models might have provided a similar comparison scenario. 

\begin{table}[!ht]
\caption{\label{tab:ml-configurations}\textsc{Modified Parameters in Different Machine Learning Classifiers. We use the default values for all the other parameters not listed in this table.}}
\centering
\addtolength{\tabcolsep}{5pt}
\begin{tabular}{|l|l|l|} 
\hline
Random Forest Classifier (\textbf{RF})                                            & Logistic Regression Classifier (\textbf{LR})                                & K-Neighbors Classifier (\textbf{KNN})  \\ 
\hline
\begin{tabular}[c]{@{}l@{}}max\_depth = 100 \\n\_estimators = 100~~\end{tabular} & \begin{tabular}[c]{@{}l@{}}solver = `liblinear’\\max\_iter = 200~~\end{tabular} & n\_neighbors = 5                       \\
\hline
\end{tabular}
\end{table}


\begin{figure}
\includegraphics[width=\textwidth]{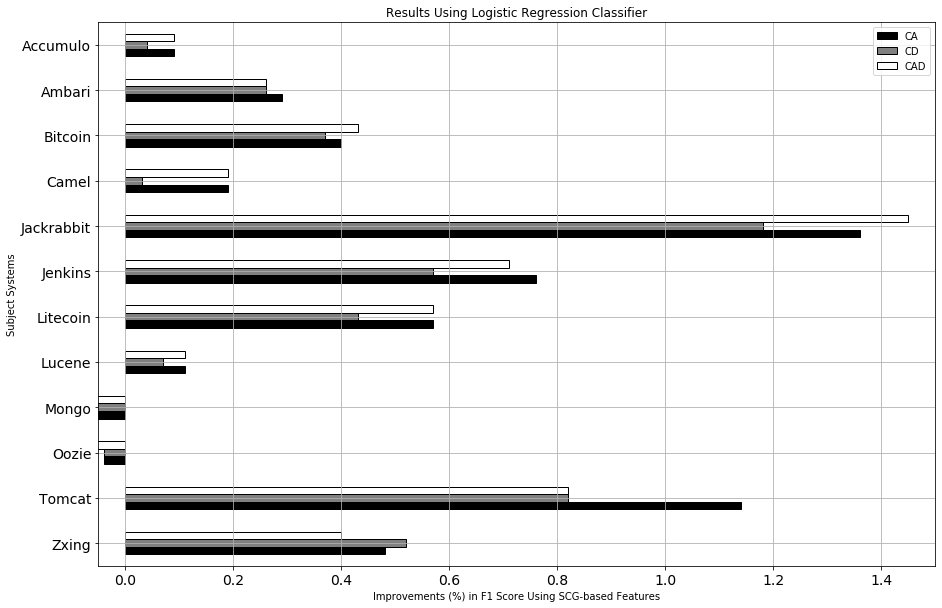}
\caption{We are comparing the improvements of F1~Scores using different Feature Combinations in the Logistic Regression Classifier. This diagram indicates the improvement of F1~scores if we use SCG-based features in combination with GitHub statistics-based features. Here, we compare the improvements for the SCG-based feature combinations, CA, CD, and CAD, to the results we obtain using only the C features. This diagram shows the improvements in F1~scores up to 140\% (e.g., Jackrabbit). We can also see a decrease in F1~Scores in Mongo and Oozie, but all the other ten subject systems increase F1~Score if we use SCG-based features combined with the C feature.}
\label{fig:ImprovementsLR}
\end{figure}

\begin{figure}
\includegraphics[width=\textwidth]{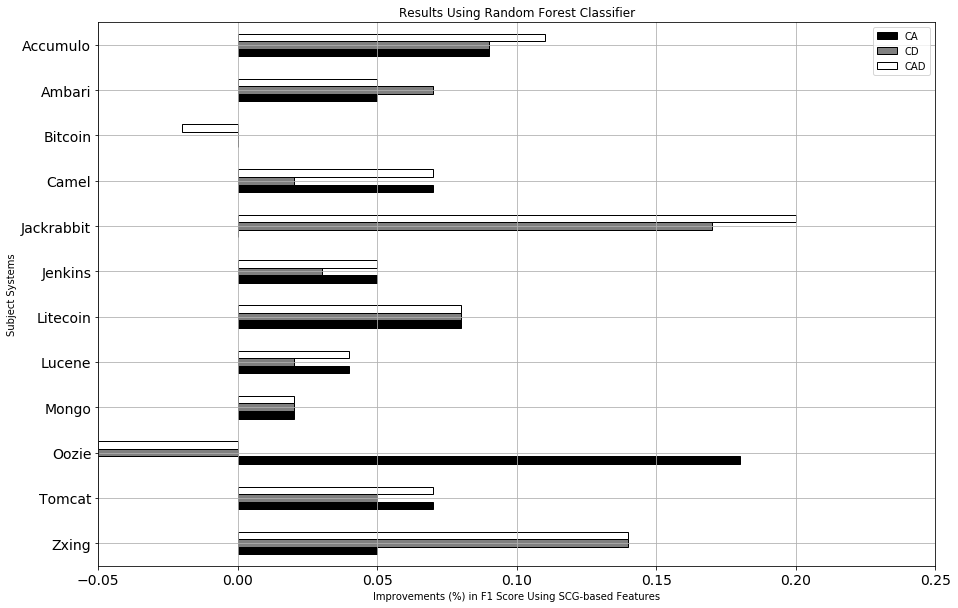}
\caption{We are comparing the improvements of F1~Scores using different Feature Combinations in the Random Forest Classifier. Here, 1.0 represents 100\%.}
\label{fig:ImprovementsRF}
\end{figure}


\begin{figure}
\includegraphics[width=\textwidth]{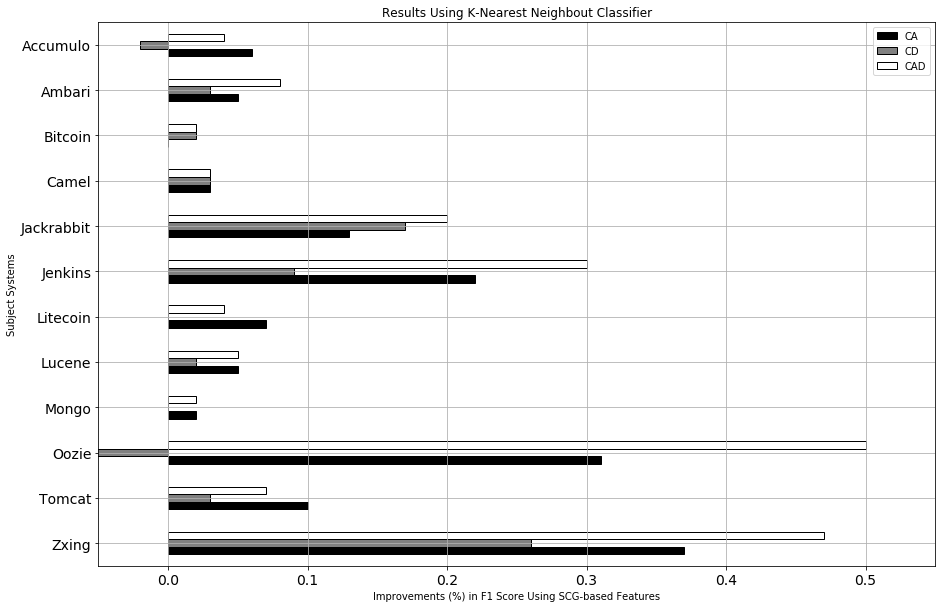}
\caption{We are comparing the improvements of F1~Scores using different Feature Combinations in the K-Nearest Neighbours Classifier. Here, 1.0 represents 100\%.}
\label{fig:ImprovementsKNN}
\end{figure}

\subsection{Evaluation Metric}
We calculate some evaluation metrics to compare the performance of buggy and non-buggy commit detection using different feature combinations. The evaluation metrics depend on the number of the detected buggy and non-buggy commit instances using ML models. The formulas for calculating the evaluation metrics are as follows.

To calculate the evaluation metrics, we need to calculate True positives (TP), True negatives (TN), False positives (FP), and False negatives. TP and TN are the numbers of commit instances correctly detected as buggy and non-buggy, respectively. On the other hand, FP and FN are the numbers of commit instances incorrectly detected as buggy and non-buggy, respectively. Therefore, a higher precision value indicates the lower FP value, which also shows higher reliability of the detected results. Similarly, a higher recall value means the lower value of FN, which also indicates the higher accuracy of the detected result. Trying to improve precision typically reduces recall and vice versa [\cite{precisionRecall}]. Thus, a detection model with higher precision with lower recall is not much accurate; on the other hand, a detection model with lower precision and higher recall is unreliable. Therefore, we calculate the F1~score, which is the harmonic mean of both the precision and recall, to compare the overall performance of our different feature combinations.

\[ Precision = \frac{TP}{TP + FP}\]
\[ Recall = \frac{TP}{TP + FN}\]
\[ F1~Score = \frac{2 \times Precision \times Recall}{Precision + Recall}\]

\section{Experimental Results}
\label{sec:results}
We generated visualization of commits from each of the 12 subject systems based on the identified labels of bug presence in those commits. The visualization of buggy and non-buggy commits is shown in Figure \ref{fig:VisContainBug}, where green dots represent the commits that do not have a bug and red dots represent the buggy commits. We have 12 subject systems and seven different feature combinations, so we generated 12$\times$7=84 different visualizations. Among them, six are shown in this figure. Investigating all the 84 visualizations (similar to the visualizations shown in this figure \ref{fig:VisContainBug}), we found the following findings. First, the addition of SCG-based features (A, D) along with GitHub based features obtained from Commit Guru implementation (C) provide better visualization compared to the only SCG-based features or only C-based features. Second, the addition of SCG-based features makes a better distinct representation of buggy commits. Finally, this visualization supports our $h_1$, as adding our graph-based features improves the separation of the commits in the buggy and non-buggy categories. 

We also investigated the results of detecting buggy and non-buggy commits using different feature combinations. The findings of detecting buggy commits using SCG-based features (A or D) and Github based Commit guru feature (C) are presented in Figures \ref{fig:ImprovementsLR}, \ref{fig:ImprovementsRF}, and \ref{fig:ImprovementsKNN}. These figures show the improvements in buggy commit detection F1~scores when combining A or D features with C features using different machine learning classification models such as Logistic Regression, Random Forest, and K-Nearest Neighbours. The improvement of F1~Scores of a buggy and non-buggy commit detection are represented using a horizontal bar chart from each of the three ML models. Figures \ref{fig:ImprovementsLR}, \ref{fig:ImprovementsRF}, and \ref{fig:ImprovementsKNN} present the improvements in F1~scores using Logistic Regression, Random Forest, and K-Nearest Neighbours classifiers, respectively. We investigated 12 subject systems and seven different feature combinations using three different machine learning classifiers. Therefore, we have 3$\times$12$\times$7=252 executions of ML models. Analyzing the results, we found CA, CD, and CAD feature combinations provide improved results in detecting buggy commits in almost all the subject systems using all the three ML models. The bar lengths in these three figures represent the per cent of improvement in F1~scores of each of the  CA, CD, and CAD features compared to the C feature.  

In Figure \ref{fig:ImprovementsLR}, we can see an improvement in F1~scores of more than 140\% using CAD feature combination in the Jackrabbit subject system. Improvements in other subject systems also remain between 40\% to 80\% in most of the subject systems. We can also see exceptions in two subject systems where the F1~scores declined when we added CA, CD, or CAD feature values, but in all the other subject systems, the SCG-based feature combination with C features improved the F1~Scores. The results using the other two machine learning classifiers in Figures \ref{fig:ImprovementsRF} \& \ref{fig:ImprovementsKNN} also provide similar results with a couple of exceptions when using the Random Forest classifier and the K-Nearest Neighbour classifier. In general, we can conclude, the inclusion of SCG based features with the C feature improves buggy and non-buggy commit detection performance in almost all of the 12 subject systems used in this study. 

SCG or Source Code Graph represents the changed code structure in a software commit patch. Different structures of a source code fragment could present different levels of complexity to the developer while changing the source code during software revision. This difference could also be a reason for a software commit to being buggy or non-buggy. Therefore, using machine learning models, the source code structure might contain strong feature values to identify buggy and non-buggy commits. The results of this study also support this scenario. The inclusion of feature values extracted from SCG helps the machine learning models to distinguish buggy and non-buggy commits more efficiently, which increases the Precision, Recall, and F1~Score in these classifiers results. However, we were also experiencing some test scenarios in Figures \ref{fig:ImprovementsLR}, \ref{fig:ImprovementsRF}, and \ref{fig:ImprovementsKNN}, where the F1~Score decreases using a combination of SCG-based features with Commit Guru features. We can see such declining cases in two subject systems (e.g., Mongo and Oozie) in Figure \ref{fig:ImprovementsLR} using the Logistic Regression classifier. Using the Random Forest classifier in Figure \ref{fig:ImprovementsRF}, we can see such a scenario resulting from Bitcoin and Oozie subject systems. Similarly, in Figure \ref{fig:ImprovementsKNN}, we can notice a similar decrease in F1~Score in the Accumulo and Oozie subject systems.

We can also notice that the Oozie subject system contains the smallest number of data instances than all the other subject systems. It is also the common subject system in the results of all the machine learning classifiers that show a decrease of F1~Scores using C, A, and D feature combinations. Thus, we can say that there are fewer data instances in Oozie, which can not perform enough training of the machine learning models to distinguish between buggy and non-buggy commits. Although using the subject system Mongo, the most extensive subject system in this investigation, also shows a decrease in F1 Scores in Figure \ref{fig:ImprovementsLR} when using the Logistic Regression classifier. Still, in Figures \ref{fig:ImprovementsRF} and \ref{fig:ImprovementsKNN}, the other two machine learning classifiers (RF and KNN) show increases in F1 Scores in the results of Mongo using SCG-based features. The internal classification mechanism of these ML classifiers might be responsible for such performance differences. We plan to investigate these differences among ML classifiers in the future study. The other two subject systems, which show a decrease in F1~Scores, are Accumulo and Bitcoin, but we can not see a reduction in all three cases (CA, CD, and CAD) of feature combinations and classifiers. F1~Score decreases only in the result of Accumulo and Bitcoin when we use CD features with KNN classifier and CAD features with Random Forest classifier, respectively.

In summary, the number of subject systems when the use of all the SCG-based feature combinations decreases the F1~Score is only 5.56\% (2 out of 36) of the total number of tested scenarios (12 $\times$ 3) in 12 subject systems using three machine learning classifiers. Therefore, 94.44 \% of the test cases of 12 subject systems in all the three machine learning models of this investigation show increase in F1~Scores by combining SCG based features and Commit Guru features. In the future, we plan to investigate whether the source code structure or the number of available data instances to train and test machine learning models in these subject systems (e.g., Accumulo, Bitcoin, Mongo, and Oozie) is responsible for reducing the F1~Score. We also plan to add additional subject systems to continue this study in the future.

We also performed the Wilcoxon Signed-Rank Test  [\cite{wilcoxon-signed-rank-test, wilcoxon-signed-rank-test-rosner}] to see the statistical significance of the improvements in F1~Scores using different feature combinations. As we have 12 subject systems, we obtained 12 F1~Scores for each of the seven feature combinations. We wanted to verify whether the improvement of F1~Scores with the addition of SCG-based features is statistically significant. In Figures \ref{fig:ImprovementsLR}, \ref{fig:ImprovementsRF}, and \ref{fig:ImprovementsKNN}, F1~Scores obtained by using CAD, CA, CD feature combinations provide improved results than F1~Scores obtained by using only C. Therefore, we did the Wilcoxon Signed-Rank Test using the F1~Scores of these four feature combinations. The resulting p values are in Table \ref{tab:significance-test}. We find all the p values less than 0.05, which indicates the improvement of results is statistically significant. To summarize our statistically significant test result, we can say that using all the three ML models, CA, CD, and CAD features provides a better result than the C features in all the subject systems using all the three machine learning models of this study.  

\begin{table}[ht]
\caption{\label{tab:significance-test}\textsc{p-Values in Wilcoxon Signed Rank Test. As all the p-Values are below 0.05, we get improved results in all the combinations of A and D features with C feature}}
\centering
\addtolength{\tabcolsep}{10pt}
\begin{tabular}{|c|c|c|} 
\hline
\textbf{ML Algorithm}                                                                    & \textbf{Comparing Features} & \textbf{p-Value}  \\ 
\hline
\multirow{3}{*}{\begin{tabular}[c]{@{}c@{}}Random Forest \\Classifier\end{tabular}}      & C vs CA                     & 0.005             \\ 
\cline{2-3}
                                                                                         & C vs CD                     & 0.049             \\ 
\cline{2-3}
                                                                                         & C vs CAD                    & 0.006             \\ 
\hline
\multirow{3}{*}{\begin{tabular}[c]{@{}c@{}}Logistic Regression\\Classifier\end{tabular}} & C vs CA                     & 0.005             \\ 
\cline{2-3}
                                                                                         & C vs CD                     & 0.010             \\ 
\cline{2-3}
                                                                                         & C vs CAD                    & 0.007             \\ 
\hline
\multirow{3}{*}{\begin{tabular}[c]{@{}c@{}}K-Nearest Neighbour\\Classifier\end{tabular}} & C vs CA                     & 0.003             \\ 
\cline{2-3}
                                                                                         & C vs CD                     & 0.039             \\ 
\cline{2-3}
                                                                                         & C vs CAD                    & 0.002             \\
\hline
\end{tabular}
\end{table}

\begin{figure}[ht]
\includegraphics[width=\textwidth]{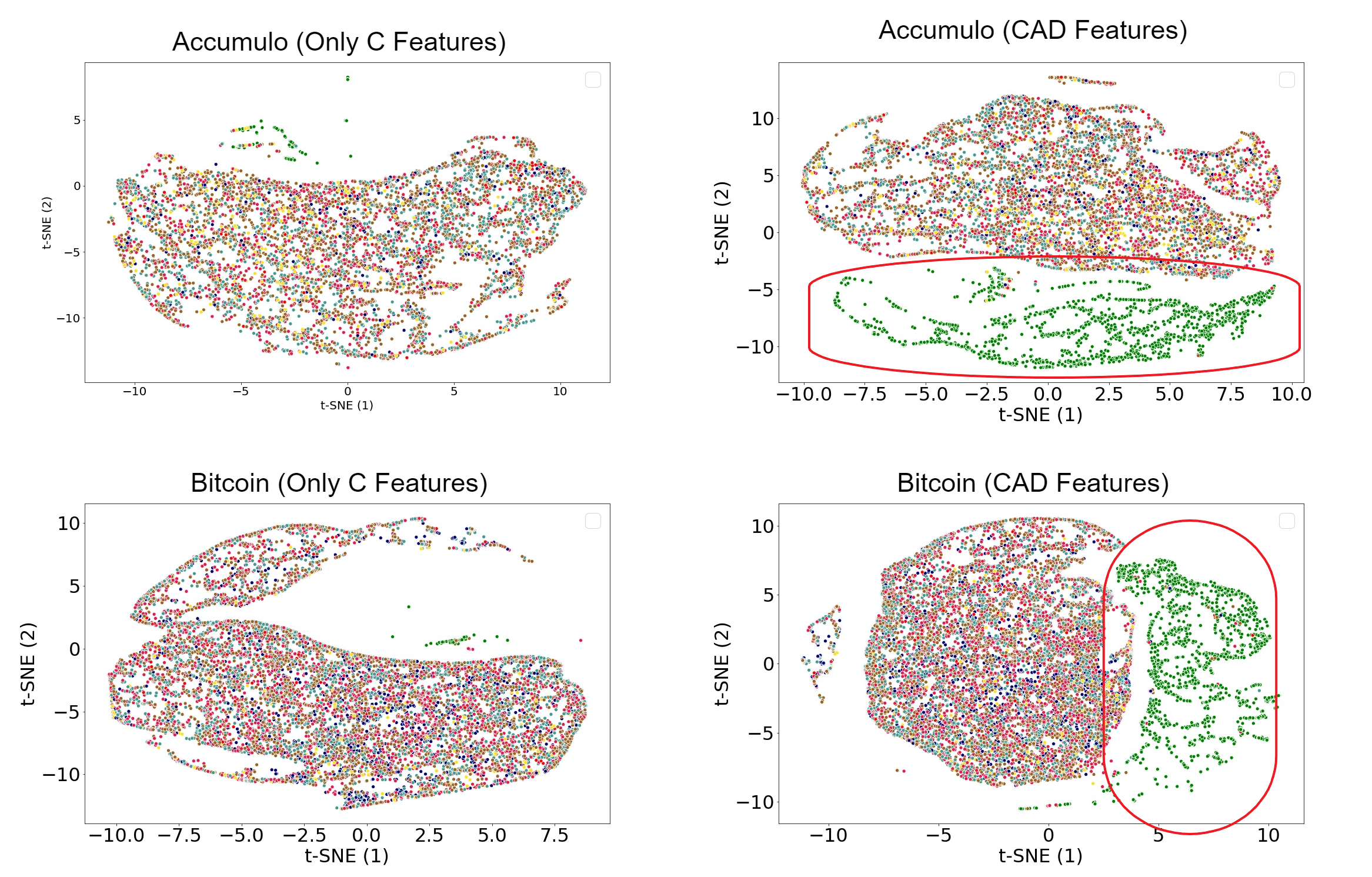}
\caption{Comparing the visualization of commit categories shows the enhanced clustering of different commit categories when we use SCG-based features in combination with the C feature (e.g., CAD on the right) compared to the visualization on the left when SCG features are not included (e.g., only C on the left). The visualization using CAD features can distinctly identify the clusters of merge commits which are represented by green dots (enclosed by a red boundary) in this image.}
\label{fig:VisClassification}
\end{figure}

The visualization of commit categories in Figure \ref{fig:VisClassification} shows when we use only the feature values C, then it is impossible to distinguish different commit categories from the generated visualization (images on the left). Using CAD (Combination of C, A, and D) feature values can produce a distinct cluster of merge commits. Although CAD feature values cannot distinguish all the seven commit categories, it indicates a separate region for the merge-commits (green dots). This visualization means merge-commits show a more distinctive structural difference in source code than other commits categories. We obtained similar visualization in all the different subject systems. In general, we can summarize that the addition of Source Code Graph (SCG) representation based on structural feature values helps produce a better visual representation of commit categories. Findings from this visualization support our hypothesis, which states that different commit categories have different coding structures. Therefore, it also answers our research questions $RQ_1$ and $RQ_2$ and suggests providing importance in the structural properly of source code while changing a codebase. We believe more investigations of these structural properties of source code could build intelligent software-driven autonomous systems and minimize the occurrences of bugs in the future. 

\section{Threats to Validity}
\label{threats-validity}
We utilized the automatically
labelled dataset and default feature values from the implementation done by [\cite{CommitGuruStudy}] and compared the visualization and classification performance using a different combination of feature values of C, A, and D. It is possible their implementation may contain incorrectly labelled commits. As in this investigation, we compare the performance improvement of visualization and ML-based models classification by using different combinations of C, A, D; we believe the impact of the incorrectly labelled dataset is insignificant in the comparison scenario of our investigation. If we had used any other data labelling methods, our comparison scenario might have provided a similar result as we got in this investigation. The analysis based on visualization is subjective to human perception. Hence a deeper investigation using controlled study can be an avenue for future research. 

We have used 12 subject systems written in C++ and Java programming languages. The use of any other programming language might have different findings. The datasets from all the 12 subject systems contain more than 246 K data instances, and we also implemented three different ML-based algorithms and seven different feature combinations. Thus, we believe our findings should be generalizable toward other software systems written in C++ or Java programming languages or other programming languages with similar architecture to C++ and Java. However, a practical verification of this assumption is required. We plan to do a similar investigation with a few other popular programming languages and software systems in the future. 

\section{Related Work}
\label{related-work}
Visualizing a software system can help to analyze the evolution of software architecture, identify the developer network, find stable software releases, and monitor software quality
trends [\cite{SoftwareVisualizationBook}]. A rich body of studies [\cite{GithruVisAnalytic, VisImpactPerformance,  ParallelEdgeSplating, VisEvo-Clocks, VisEvolvingGenealogy}] explored different ways to visualize evolving software systems for making it easily understandable to keep the consistent evolution. \citet{StructureVisualizaionAST} did a visualization of evolution patterns in C++ source code by rendering syntax matched code blocks in consecutive versions to detect the code fragments which have been changed during evolution. There are several tools for visualizing software systems [\cite{VisSoftwareTool:1, VisSoftwareTool:2, VisSoftwareTool:3}]. Among those tools, many were mainly used by software architects [\cite{VisICSE2018EVA}], who are interested in the high-level abstraction of the software system. They are not interested in coding details. \citet{VisualizingCodePatternNovice} examined code patterns by visualizing the code written by novice programmers such as first-year students. They wanted to help either instructors or students identify poor programming practices during the coding process. \citet{VisTestFault} implemented source code visualization for locating defects or faults in the source code. Studies were also done to help the maintenance activities by visualizing the software system [\cite{VisCollaborationMaintenance}]. \citet{VisPerformanceEvaluationMatrix} proposed an interactive visualization technique that allows practitioners to analyze the performance of software components along with multiple software versions at once. \citet{SDiehlSoftwareVis} identified three categories (structure, behaviour, and evolution) for using visualization techniques to explore software systems. 

There are a bunch of studies [\cite{Asaduzzaman:Bug:Inducing:USASK, Bavota:2012:RIB:2477171.2477400, Bernardi:Developer:Induce:Bug, Canfora:How:Long:Bug:Survive, Ell:Failure:Inducing:Developer:Pair, Eyolfson:2011:TDD:1985441.1985464, Kim:2006:LDT:1137983.1138027, Sliwerski:2005:HRR:1095430.1081725}] to investigate software commit operations that can induce new bug (bug inducing commit) in software systems, which is known as Just-In-Time (JIT) [\cite{Rel:2015:Yang:Deep:JIT:Defect:Prediction}] bug prediction in literature. Most of the studies for JIT bug prediction utilized source code repository-based statistics [\cite{2019:Borg:SZZUnleashed, CommitGuruRosen}] of commit operations such as the number of lines added, deleted, or modified. \citet{ReducingFeatures-BIC} treated the source code as natural language text [\cite{NaturalnessSoftware, NaturalnessBug}] and utilized Natural Language Processing (NLP) \cite{NLPHistorical} techniques to extract feature values to apply ML-based detection models. Deep learning techniques, such as Deep Belief Networks (DBN) [\cite{SemanticFBugPrediction}], have also been used to learn semantic features from the abstract syntax trees of the program. In this study, we used the structural properties of source code by representing code segment in commit patch as a graph structure. Compared to all the existing studies in literature, our study emphasized the structural attributes of the source code and reached the performance improvement in detecting bugs.     

We did not find any study that utilized the features obtained from source code visualization to identify bug presence in software systems using machine learning-based detection models. However, several studies [\cite{TBCNN, DeepSimilarity, DeepBugs, Gra:ASTLan, allamanis2018learning, PDGOptimization, SemanticFBugPrediction}] proposed different representations to preserve syntactic and semantic information in source code embedding, but they require compiled intermediate model of byte-codes [\cite{astnnICSE2019, LLVMCompiler, PreciseDataFlowAlg}]. These techniques are not directly applicable for the source code fragments in the commit patch as they are uncompilable and incomplete. \citet{astnnICSE2019} is the only study that can process incomplete and uncompilable source code for converting them numeric vector representation, which could use in ML-based models. They implemented Abstract Syntax Tree (AST) [\cite{cloneDetectionAST}] based Neural Network (ASTNN) to learn vector representation of source code fragments and applied them to classify source code and find code clones. Our work differs from their work in the final embedding of source code-based feature values. They utilized the Word2Vec [\cite{Word2Vec}] model for making numeric vector-based representation from the generalized tokens obtained from the AST of a source code fragment. 

On the other hand, we converted the generalized tokens of the source code fragment into a graph from which we extracted distinct feature values. In addition, during the processing of source code fragments by Abstract Syntax Tree (AST) based Neural Network (ASTNN), they created a separate numeric vector representation for each of the code fragments, making it challenging to adapt the technique for the commits that have multiple code fragments changes. In our investigation, we found most of the commit operations have multiple changed fragments, and we created separate graph representation for each of the fragments and then took the union of all the graphs [\cite{GraphsUnion}] in a commit to forming a single graph. The union of graphs makes it very easy to represent a single commit by visualization using graph representation and to take the attributes of the final graph to form a feature set for the commit. Since a graph structure often encodes rich information, we believe our graph-based feature will provide a more easily understandable representation of source code complexity and developers' coding style. 


\section{Conclusion \& Future Work}
\label{conclusion-future-work}
Source code in software systems experiences several changes to adapt to different requirements in its evolution \& maintenance process. Sometimes a change done to solve a problem (bug) in a software system may also be a reason for introducing new issues (bug introducing) in that system, requiring more subsequent changes and attention to eliminate the introduced bug. The key objective of this investigation was to produce a graph representation based on the programming styles of software developers. The attribute values from these representations are used in detecting and visualizing commits during software revisions. We introduced Source Code Graph (SCG) based feature values and validated their use to visualize and detect buggy and non-buggy commits from 12 C++ and Java-based software systems. All the software systems used in this study contain more than 246 K commit instances.  This study also investigated the use of SCG-based feature values to visualize the buggy and non-buggy commits and seven commit categories. The visualizations we implemented show that the addition of SCG-based features can distinctly identify the merge commits and improve the detection and visualization of other commit categories. As SCG-based features mainly represent the style of writing source code, this study summarizes the importance of developers' coding style or patterns. These findings motivate us to do more experiments on these source code patterns in future. The findings of this research can help us to produce guidelines for software developers to avoid coding structures that are more likely to induce bugs in software systems.  Avoiding an inconsistent/ risky coding structure on a Maintenance Phases of Software Engineering might help the developer avoid unstable/ buggy software evolution. 

Although all the visualizations of bug presence and commit classification in this study have been improved by adding  Source Code Graph (SCG) based features, it still can not entirely separate distinct regions among the commit categories. We plan to do more investigations in future on the extraction and encoding of SCG-based features to identify different commit categories effectively (e.g., automatically identifying 1-Merge, 2-Corrective, 3-Preventive, and other commits categories as presented in Table \ref{tab:vis-commit-classification}). In the future, we also plan to extend this study to find and represent specific coding styles and patterns most common in buggy software revisions. The continuation of this study in more detailed investigations can help to propose the correct coding style or guideline for effective and bug-free software evolution \& maintenance. 

In this study, we show a novel approach to encode source code structure into a graph representation and then utilize the attributes of that graph to identify buggy and non-buggy software commits. We can also investigate the techniques applied for encoding and using structural properties of source code in some related studies, such as detecting replicated source codes in the code repositories of different software systems [\cite{CodeCloneWeb:IRO, astnnICSE2019, cloneDetectionAST}], and implementing automatic identification of software defects (e.g., faults or bugs) [\cite{SoftwareTestingHybrid:IRO}]. Therefore, we plan to investigate these new research directions in future studies.

\begin{acknowledgements}
This work is supported by the Natural Sciences and Engineering Research Council of Canada (NSERC) and by two Canada First Research Excellence Fund (CFREF) grants coordinated by Global Institute for Food Security (GIFS) and Global Institute for Water Security (GIWS).
\end{acknowledgements}

%
%

\bibliographystyle{spbasic}      
\bibliography{citations}   

%
%

\end{document}